\documentclass[11pt,letterpaper,twocolumn]{article}

\usepackage[margin=0.75in]{geometry}
\usepackage[T1]{fontenc}
\usepackage[utf8]{inputenc}
\usepackage{microtype}
\usepackage{graphicx}
\usepackage{amsmath,amssymb}
\usepackage{booktabs}
\usepackage{tabularx}
\usepackage{array}
\usepackage{listings}
\usepackage{hyperref}
\usepackage{cleveref}
\usepackage{xcolor}
\usepackage{url}
\usepackage{caption}
\usepackage{enumitem}

\hypersetup{
  colorlinks=true,
  linkcolor=blue!50!black,
  citecolor=blue!50!black,
  urlcolor=blue!50!black
}

\title{memorywire: A Vendor-Neutral Wire Format for Agent Memory Operations}
\author{
  Thamilvendhan Munirathinam\thanks{\texttt{mthamil107@gmail.com}}\\
  Independent Researcher\\
  \small{Repository: \url{https://github.com/mthamil107/memorywire}}
}
\date{}

\begin{document}
\maketitle

\begin{abstract}
Agent-memory frameworks --- mem0, Letta/MemGPT, Cognee, Zep/Graphiti, MemoryOS, MemTensor --- each ship their own SDK, storage layout, and operational vocabulary. There is no shared wire format: every integration is bespoke, every migration rebuilds memory from scratch, and no framework ships a governance surface that lets a human review writes before they enter long-term storage. We present \texttt{memorywire}\footnote{Originally drafted as ``Agent Memory Protocol (AMP)'' in May 2026; renamed to \texttt{memorywire} before launch to avoid collision with an unrelated, prior project of the same name (\url{https://github.com/akshayaggarwal99/amp}, created Dec 2025). See \texttt{docs/PRIOR-WORK.md} in the repository for the long version.}, a JSON-Schema 2020-12 wire format for five memory operations (\texttt{remember}, \texttt{recall}, \texttt{forget}, \texttt{merge}, \texttt{expire}) over four memory types (semantic, episodic, procedural, emotional), with a \texttt{MemoryStore} interface, a fan-out router, and an optional HITL governance channel. We describe an open-source reference implementation with five backend adapters (sqlite-vec, mem0, Letta, Cognee, pgvector); a microbenchmark on a 100-fact / 50-query labelled corpus (42 with non-empty gold ids + 8 no-match probes) achieving recall@5 = 1.000 on the 42 gold-id queries with ingest p50 = 37.8\,ms and recall p50 = 40.6\,ms; an adversarial-fusion experiment showing Reciprocal Rank Fusion holds recall@5 = 1.000 across a 1-of-$N$ rank-0 injection sweep ($K \in \{0,5,\dots,50\}$) where \texttt{max} fusion collapses to 0.500 with 80\% leak at $K \geq 5$; and a 16-scenario cross-adapter conformance suite passing 68 of 80 cells with zero failures. The contribution is not a new algorithm; it is a packaging of established components (RRF, FSMs, STM/LTM consolidation, diff-and-approve workflows) into a venue-neutral protocol with an empirically validated reference, positioned to compose with the Model Context Protocol rather than compete with it. We further show that \texttt{memorywire}'s provenance field is the strongest lever for \emph{recovering} a poisoned store once it is compromised, evaluated with our companion benchmark PurgeBench~\cite{purgebench}.
\end{abstract}

\section{Introduction}
\label{sec:intro}

\subsection{The problem: islanded memory frameworks}
\label{sec:intro-problem}

Agent runtimes that maintain memory across sessions are now a category. Open-source frameworks include mem0, Letta (formerly MemGPT), Cognee, Zep/Graphiti, MemoryOS, and MemTensor MemOS; closed commercial offerings include Oracle's AI Agent Memory and the memory layers shipped inside major hosted-agent platforms. What the category has not produced is a shared wire format. Each framework defines its own SDK surface, JSON shape for memory records, embedding-provider integration, taxonomy (or absence of taxonomy) for memory types, and implicit lifecycle for record creation and deletion. The heterogeneity is non-trivial to bridge: mem0 stores records under a \texttt{memories[]} list keyed by \texttt{user\_id} with a heterogeneous \texttt{created\_at} representation; Letta stores archival memory keyed by \texttt{agent\_id} and exposes a \texttt{tags} list as the only structured-metadata sink; Cognee mints internal \texttt{data\_id} UUIDs that are not surfaced through its public \texttt{add} API, making per-record deletion impossible from outside the pipeline; sqlite-vec stores tables keyed by a stable ULID-shaped string; pgvector exposes records through an application-chosen SQL schema. Re-platforming an agent from one framework to another therefore requires a bespoke migrator and field-level losses where the source framework encodes more state than the target's data model holds.

The same heterogeneity means there is no shared \emph{governance} surface. Each framework provides a write API and a read API; none mediate the write with a ``diff against current state, present to a human, commit only on approval'' workflow. Production governance work such as Taheri's \emph{Governed Memory}~\cite{taheri2026governed} enforces write policy automatically across autonomous agents rather than through such a human-approval gate, so this human-in-the-loop workflow has no production implementation an off-the-shelf agent can drop in. Operators who want auditability over what enters long-term memory must build it themselves and accept that the framework can bypass them.

This is the gap memorywire addresses. It is not ``we need a better retrieval algorithm'' --- the algorithms in the category (vector search, hybrid lexical-semantic RRF fusion, graph hop boosts, FSM-encoded procedures, STM/LTM consolidation) are well understood. It is ``we need a shared protocol so any client can talk to any backend, any agent can carry its memory across runtimes, and any write can be diffed and approved.'' Structurally it is the gap MCP closed for \emph{tool use}, applied to \emph{memory}.

\subsection{Contributions}
\label{sec:intro-contributions}

This paper makes five contributions:

\begin{itemize}[leftmargin=*]
  \item \textbf{C1.} A wire format for five memory operations over four memory types, expressed as JSON Schema 2020-12~\cite{jsonschema_2020_12} (\texttt{docs/spec/v0.md}). The operations are \texttt{remember}, \texttt{recall}, \texttt{forget}, \texttt{merge}, \texttt{expire}; the types are \texttt{semantic}, \texttt{episodic}, \texttt{procedural}, \texttt{emotional}. The schemas are vendor-neutral, transport-agnostic, and explicitly versioned with a breaking-change policy through v0.5.
  \item \textbf{C2.} A reference implementation in Python 3.11+ with five production-backend adapters (sqlite-vec, mem0, Letta, Cognee, pgvector), all implementing a single \texttt{MemoryStore} Protocol. The reference includes a memory router that fans operations across $N$ stores in parallel and fuses recall results via Reciprocal Rank Fusion ($k=60$)~\cite{cormack2009rrf} with an optional one-hop graph boost, plus a tolerant partial-failure model where a single rogue or unavailable backend cannot crash the operation.
  \item \textbf{C3.} A governance UI implementing a human-in-the-loop diff-and-approve workflow over \texttt{remember}, \texttt{forget}, and \texttt{merge}. Writes flagged \texttt{approval\_required} are staged behind a \texttt{PENDING\_APPROVAL\_DELETED\_AT = -1} sentinel and remain invisible to \texttt{recall} until a reviewer commits or rejects them through the UI. The same audit log is the single source of truth for all governance and mutation events.
  \item \textbf{C4.} An empirical evaluation comprising (a)~a microbenchmark on 100 hand-authored facts $\times$ 50 labelled queries against a real sentence-transformer embedder, (b)~an adversarial-fusion experiment that sweeps a 1-of-$N$ rank-0 injection attack across three fusion algorithms (RRF, MAX, weighted), and (c)~a cross-adapter conformance suite of 16 protocol-invariant scenarios run against all five shipped adapters (68 PASS / 12 SKIP / 0 FAIL out of 80 cells).
  \item \textbf{C5.} A six-adversary threat model with line-level mitigation citations into the reference implementation, plus an open-data artifact (\texttt{docs/adversarial-results.\{rrf,max,weighted\}.json}, the labelled microbench corpus, the conformance scenario list) sufficient to reproduce every empirical claim without re-running paid evaluators.
\end{itemize}

\subsection{Limitations, front-loaded}
\label{sec:intro-limitations}

We state up front what this paper is \emph{not}. It is not a new algorithm: RRF is from Cormack et al.~\cite{cormack2009rrf}, the four-type taxonomy is from Tulving~\cite{tulving1972episodic} and Squire~\cite{squire1992declarative}, procedural FSMs use \texttt{pytransitions}, and STM/LTM consolidation predates the frameworks we cite --- the contribution is composition and standardization, not new substrate. It is not an LLM contribution: we do not train models or propose model-dependent heuristics, and we report only \emph{preliminary} LongMemEval~\cite{wu2024longmemeval} and LoCoMo~\cite{maharana2024locomo} numbers (\Cref{sec:eval-mem-benchmarks}); the full 5-seed $\times$ 200-question run and BEAM are v2 replacement deliverables (\Cref{sec:future}). And it is not a stabilization paper: memorywire v0 reserves the right to break wire-format compatibility through v0.5 (\Cref{sec:spec-versioning}); the IETF Internet-Draft and MCP-WG extension paths in \Cref{sec:related} and \Cref{sec:future} are intent, not commitment.

\subsection{Paper roadmap}
\label{sec:intro-roadmap}

\Cref{sec:related} places memorywire against prior work in agent memory frameworks, cross-vendor protocols (particularly MCP), and prior production work on memory governance. \Cref{sec:spec} specifies the wire format: operations, types, the \texttt{MemoryStore} Protocol, router semantics, and the governance channel. \Cref{sec:impl} describes the reference implementation including the five backend adapters, the procedural-memory FSM backend, the STM$\leftrightarrow$LTM transformer, and the governance UI. \Cref{sec:eval} reports the empirical evaluation. \Cref{sec:threats} is the threat model. \Cref{sec:mcp} details the relationship to MCP. \Cref{sec:future} and \Cref{sec:conclusion} lay out future work and the bet we are making.

\section{Background and Related Work}
\label{sec:related}

\subsection{Memory in LLM agents}

The first wave of LLM-agent frameworks treated memory as a side-effect of the conversation log; the next externalized it into a vector store (``RAG over your own logs''); the current wave, beginning roughly with MemGPT~\cite{packer2023memgpt} and continuing through mem0~\cite{chhikara2025mem0}, Cognee, Zep/Graphiti, MemoryOS, and MemTensor, separates memory into named tiers (short-term working, long-term semantic, episodic, procedural) with explicit lifecycle operations (consolidate, expire, merge).

The frameworks differ in which tiers they emphasize. MemGPT/Letta exposes a hierarchical ``core memory'' + ``archival memory'' model with explicit \texttt{core\_memory\_append} / \texttt{archival\_memory\_insert} tools. Mem0 emphasizes user-scoped facts with an internal LLM-assisted dedupe pass. Cognee constructs a knowledge graph and exposes traversal as the primary recall primitive. Zep/Graphiti adds temporal edges over a graph. MemoryOS positions itself as an OS-shaped memory layer; MemTensor MemOS extends this with multi-tenant scoping.

None of these frameworks publishes a wire format that another framework adopts. The closest thing to a shared shape is mem0's request structure, informally copied by smaller projects, but mem0's API is the surface of one library, not a specification --- it can change between releases.

\subsection{Cross-vendor protocols: MCP as template}

The Model Context Protocol~\cite{mcp_spec_2025} is the most successful recent cross-vendor protocol in the agent space. MCP is a JSON-RPC protocol over stdio or HTTP standardizing how an agent runtime exchanges \emph{context} with a server process; its public surface is five primitives --- tools, resources, prompts, sampling, and roots/elicitation --- and it has been adopted by Anthropic, OpenAI, Google's Gemini ecosystem, Cloudflare, and many smaller server implementations.

MCP does not define a memory primitive. A backend can be \emph{wrapped} in MCP --- exposing \texttt{remember} / \texttt{recall} / \texttt{forget} as three tools is a fifteen-minute integration --- but the wrapping is lossy: operations flatten into opaque tools, the type taxonomy collapses into a string parameter, and the governance channel sits entirely outside MCP. We discuss the memorywire--MCP relationship at length in \Cref{sec:mcp}; the short version is that memorywire is what MCP would be if MCP had a memory primitive, built as a standalone spec so the design can stabilize without blocking on MCP-WG governance, with intent to propose it as an MCP extension at v0.5. The standalone-first, propose-upstream-later pattern follows Cloudflare's Web Bot Auth precedent~\cite{meunier2026webbotauth}.

\subsection{Reciprocal Rank Fusion}
\label{sec:related-rrf}

The memory router's default fusion algorithm is Reciprocal Rank Fusion (RRF) from Cormack, Clarke, and Buettcher~\cite{cormack2009rrf}, who showed that RRF outperforms Condorcet and individual rank-learning methods on TREC tasks. The formula is
\[
  \mathrm{score}(d) = \sum_{i \in S} \frac{1}{k + r_i(d)}
\]
where $S$ is the set of ranked lists, $r_i(d)$ is the rank of document $d$ in list $i$, and $k$ is a smoothing constant (60 in the original paper; memorywire uses the same value as a sensible default). RRF has two properties that matter for our setting: (a) it is \emph{score-independent} --- the input lists' raw scores do not enter the fusion, only their ranks do; (b) it is \emph{additive across stores} --- the contribution of an item to the fused score is the sum of its per-list contributions. These two properties together give RRF a robustness against rogue stores that score-sensitive fusion methods do not have. We exploit this in \Cref{sec:eval-adversarial}.

\subsection{Human-memory taxonomy}

memorywire's four memory types --- \texttt{semantic}, \texttt{episodic}, \texttt{procedural}, \texttt{emotional} --- are drawn from the human-memory taxonomy in cognitive-science literature. Tulving~\cite{tulving1972episodic} introduced the semantic / episodic distinction; Squire~\cite{squire1992declarative} elaborated the declarative / nondeclarative hierarchy in which procedural memory sits as a nondeclarative subtype. The \texttt{emotional} type is included as a first-class tag rather than buried in metadata because affective associations are operationally distinct in agent workflows (they often trigger or suppress specific tools) and because the alternative --- encoding them as semantic facts with a sentiment field --- collapses the distinction at the wire-format layer where it matters most.

We do not claim that the four-type taxonomy is the \emph{correct} one in any deep sense. We claim only that it is the taxonomy most often referenced by the existing memory frameworks (Letta, MemoryOS, MemTensor all cite Tulving/Squire) and that pinning it at the wire-format layer is more useful than leaving the choice to each backend.

\subsection{Human-in-the-loop approval for agent actions}

The governance channel in memorywire implements a human-in-the-loop diff-and-approve workflow: when an agent proposes to write a memory, the system computes a structured diff between the proposed write and the current state, presents the diff to a human reviewer, and commits the write only on approval. This contrasts with production governance for autonomous agents such as Taheri's \emph{Governed Memory}~\cite{taheri2026governed}, which enforces write policy automatically rather than routing writes through a human gate. The workflow generalizes to any state mutation; memorywire applies it to \texttt{remember}, \texttt{forget}, and \texttt{merge}, and excludes \texttt{recall} and \texttt{expire} from the default approval surface (with \texttt{recall} flagged for v0.2 reconsideration; see \Cref{sec:threats}).

Diff-and-approve as a review discipline is not itself novel---it mirrors code review and change-management gates. What is new here is its standardization at the wire-format layer: memorywire defines a \texttt{governance} JSON schema for the diff-and-approve message and ships a reference UI that any backend adapter inherits transparently. An agent calling \texttt{remember(content="\ldots", approval\_required=true)} gets the governance flow regardless of which of the five backends actually stores the row.

\section{The memorywire Wire Format}
\label{sec:spec}

\subsection{Design goals}
\label{sec:spec-goals}

Four explicit goals shaped the v0 surface.

\textbf{Small surface.} Five operations, four types, one optional governance channel. The operations are the smallest set we found that cover every workflow exhibited by the surveyed frameworks: \texttt{remember} (write), \texttt{recall} (read), \texttt{forget} (delete), \texttt{merge} (deduplicate), \texttt{expire} (apply a TTL policy). Adding more operations is straightforward; we resisted doing so in v0 because every new operation is a new schema to keep stable through the v1.0 freeze.

\textbf{Schema-pinned.} Every operation has a JSON Schema 2020-12 file in \texttt{src/memorywire/schemas/operations/}. Validation runs at the boundary; backends accept already-parsed \texttt{pydantic} models that mirror the schemas exactly. The schemas are the source of truth, the models are the convenience layer.

\textbf{Stateless.} Each operation request carries the \texttt{agent\_id} (and optional \texttt{user\_id}) it operates under; the router does not keep per-session state. This is what makes the protocol portable across transports --- REST today, JSON-RPC tomorrow, any future cross-language port without redesigning the wire format.

\textbf{Async-first.} The \texttt{MemoryStore} Protocol's methods are \texttt{async def}. The router uses \texttt{asyncio.gather} with \texttt{return\_exceptions=True} for fan-out, so per-store failures do not cascade. Sync convenience wrappers are layered above through \texttt{asgiref} for callers who want them; the canonical path is async.

\subsection{Operations}
\label{sec:spec-ops}

Each operation has a request schema and a response schema. We summarize the request side here; full schemas live in \texttt{src/memorywire/schemas/operations/} and worked examples in \texttt{docs/spec/examples/}.

\textbf{\texttt{remember}} writes a new memory. Required: \texttt{agent\_id}, \texttt{type} (one of four type tags), \texttt{content}. Optional: \texttt{user\_id}, \texttt{metadata} (free-form JSON), \texttt{confidence} (float in $[0,1]$, default 1.0), \texttt{source}, \texttt{expires\_at} (Unix epoch ms), \texttt{approval\_required} (bool, default false). \texttt{type} is pinned at write time so the router can route procedural writes to the FSM backend; \texttt{confidence} is first-class because every surveyed framework either has a confidence equivalent or asks for one.

\textbf{\texttt{recall}} reads memories. Required: \texttt{agent\_id}, \texttt{query}. Optional: \texttt{k} (1--1000, default 5), \texttt{types} (filter subset), \texttt{hops} (0--3, default 0), \texttt{fusion} (\texttt{rrf}/\texttt{max}/\texttt{weighted}, default \texttt{rrf}), \texttt{filter}, \texttt{fresher\_than\_days}. \texttt{fusion} is exposed because the right choice depends on the operator's trust model (see \Cref{sec:eval-adversarial}).

\textbf{\texttt{forget}} deletes memories. Required: \texttt{agent\_id}, plus at least one of \texttt{ids} or \texttt{filter} (the spec rejects requests with neither --- no-scope-mass-delete protection). Optional: \texttt{hard\_delete} (bool, default false), \texttt{reason}. Default soft delete preserves the audit trail; hard delete is available for GDPR takedowns.

\textbf{\texttt{merge}} collapses duplicates. Required: \texttt{agent\_id}, \texttt{canonical}, \texttt{duplicates}. Optional: \texttt{strategy} (one of \texttt{keep\_canonical} / \texttt{merge\_content} / \texttt{keep\_highest\_confidence}, default \texttt{keep\_canonical}). Pinning the strategies at the wire-format layer means portable agent code does not need to know which framework's internal dedupe will fire.

\textbf{\texttt{expire}} applies a TTL policy. Required: \texttt{agent\_id}. Optional: \texttt{policy} (ANDed \texttt{older\_than\_days}, \texttt{type}, \texttt{confidence\_below}, \texttt{no\_recall\_in\_days} predicates) and \texttt{action} (\texttt{forget}/\texttt{archive}/\texttt{demote}, default \texttt{forget}). Every framework has some notion of ``delete old episodics'' or ``down-rank cold facts,'' each expressed differently; memorywire pins the predicate DSL and action vocabulary.

\subsection{Memory types}
\label{sec:spec-types}

The four memory types are drawn from \Cref{sec:related} above. Their definitions in memorywire v0 are summarized in \Cref{tab:types}.

\begin{table}[t]
\centering
\caption{The four memorywire memory types with example content.}
\label{tab:types}
\small
\begin{tabularx}{\columnwidth}{lX}
\toprule
\textbf{Type} & \textbf{Definition / Example} \\
\midrule
\texttt{semantic}   & Declarative facts. \emph{``Alice is allergic to peanuts''} \\
\texttt{episodic}   & Past events with time/place. \emph{``On 2026-05-20 Alice told me she was nervous about her flight''} \\
\texttt{procedural} & How-to / FSM-encoded procedures. \emph{State machine for ``book a flight''} \\
\texttt{emotional}  & Affective associations. \emph{``Alice expressed anxiety when discussing flights''} \\
\bottomrule
\end{tabularx}
\end{table}

Backends may store all four types in a single table with a \texttt{type} column or shard by type. The wire format does not constrain storage strategy; it does require that the \texttt{type} tag round-trips through \texttt{remember} $\rightarrow$ \texttt{recall} losslessly.

\subsection{The \texttt{MemoryStore} Protocol and capability declarations}
\label{sec:spec-protocol}

A backend adopts memorywire by implementing the \texttt{MemoryStore} Protocol (Python; the same shape ports trivially to other languages):

\begin{lstlisting}[language=Python]
class MemoryStore(Protocol):
    async def remember(self, req: RememberRequest) -> RememberResponse: ...
    async def recall(self,   req: RecallRequest)   -> RecallResponse:   ...
    async def forget(self,   req: ForgetRequest)   -> ForgetResponse:   ...
    async def merge(self,    req: MergeRequest)    -> MergeResponse:    ...
    async def expire(self,   req: ExpireRequest)   -> ExpireResponse:   ...
    async def health(self) -> dict[str, Any]: ...
    @property
    def capabilities(self) -> set[str]: ...
\end{lstlisting}

\texttt{capabilities} declares the optional surface a backend supports --- e.g. \texttt{\{"semantic", "episodic", "vector", "fts", "graph", "procedural", "recall\_tracking"\}}. The router consults \texttt{capabilities} to skip stores that do not support a given operation. This is what lets a single router compose a vector-only adapter with a graph-only adapter without the caller needing to know which fields each backend honors.

\subsection{The memory router}
\label{sec:spec-router}

The router (\texttt{src/memorywire/router.py}) is itself a \texttt{MemoryStore}. It is constructed over $N$ child stores and fans each operation across them. For \texttt{remember}, the default policy is \texttt{fan\_out}: write to every child store. For \texttt{recall}, every store is queried in parallel and the results are fused. For \texttt{forget}, \texttt{merge}, \texttt{expire}, fan-out aggregates per-store responses with per-store errors preserved.

\textbf{Fusion math.} Three fusion algorithms are exposed:

\begin{itemize}[leftmargin=*]
  \item \textbf{RRF (default).} Per the formula in \Cref{sec:related-rrf}, with $k=60$:
    \[
      \mathrm{score}(d) = \sum_{i \in S} \frac{1}{60 + r_i(d)}
    \]
    The item's own per-list score is ignored; only its rank in each list contributes (\texttt{src/memorywire/router.py:415--444}).
  \item \textbf{MAX.} $\mathrm{score}(d) = \max_{i \in S} s_i(d)$ where $s_i(d)$ is the raw score in list $i$. Score-sensitive.
  \item \textbf{WEIGHTED.} $\mathrm{score}(d) = \sum_{i \in S} w_i \cdot s_i(d)$ with per-store weights $w_i$ from router config. Score-sensitive.
\end{itemize}

\textbf{Graph-hop boost.} When a recall request sets \texttt{hops > 0} and at least one child store declares \texttt{graph} capability, the router applies a multiplicative boost to fused items whose neighbours are also in the fused set:
\[
  \mathrm{final}(d) = \mathrm{rrf}(d) \cdot \left(1 + \frac{0.1}{1 + \mathrm{hop\_dist}(d)}\right)
\]
The boost only fires for items already in the fused set; new neighbours are not introduced (per \texttt{src/memorywire/router.py:482--501}). This avoids amplifying a malicious graph store's ability to inject content (see \Cref{sec:threats-poisoned}).

\textbf{Partial-failure model.} Fan-out uses \texttt{asyncio.gather(\ldots, return\_exceptions=True)} (\texttt{src/memorywire/router.py:261--264, 353--356}). A backend that throws or times out is logged but does not abort the operation; the router proceeds with results from the surviving stores. This is what makes a five-adapter router tolerant in practice --- a transient mem0 outage does not blank the recall.

\subsection{The governance channel}
\label{sec:spec-governance}

When the host wires in a \texttt{GovernanceClient}, \texttt{remember} / \texttt{forget} / \texttt{merge} operations carrying \texttt{approval\_required: true} (or matching a policy rule) are routed through governance before commit. The governance review request is a JSON object carrying \texttt{operation} (one of \texttt{remember}/\texttt{forget}/\texttt{merge}), \texttt{agent\_id}, the original \texttt{request}, a structured \texttt{diff} against current state, and an optional \texttt{reasoning} string. The governance response carries \texttt{approved} (bool), \texttt{reviewer} (identity), \texttt{reviewed\_at} (Unix epoch ms), and an optional \texttt{reason}. The reference UI layers an opt-in approval-learning loop on top: track which patterns the reviewer always approves / rejects and auto-allow after $N$ consistent decisions, with every fired decision still journaled to the audit log.

Excluding \texttt{recall} from the default governance surface is a v0 trade-off. Recall budgets and read-side approval are tracked for v0.2 (\Cref{sec:future}); until then, the audit log captures every \texttt{recall} call (\texttt{operation = 'recall'}) so high-rate exfiltration is observable post-hoc, if not preventable.

\subsection{Spec versioning and backwards-compatibility rules}
\label{sec:spec-versioning}

memorywire v0 is a draft. Breaking changes are allowed through v0.5; v0.5 stabilizes; v1.0 freezes. The rules are: new \emph{optional} fields are allowed at any time; new \emph{required} fields require a minor version bump with a deprecation period; \emph{removed} fields require a major version bump only; new memory types require a minor version bump. The intent is to publish memorywire v0.5 as an IETF Internet-Draft (\texttt{draft-<name>-memorywire-00}) and, in parallel, propose adoption as an MCP extension (\texttt{mcp.memory.v0} or similar; see \Cref{sec:mcp}). Both outcomes are acceptable; neither is committed.

\section{Reference Implementation}
\label{sec:impl}

The reference implementation lives in \texttt{src/memorywire/} (the SDK and adapters) and \texttt{ui/} (the governance UI). It is Python 3.11+, Apache-2.0 licensed for the protocol and reference; the Pro UI ships under the Fair Source License (FSL). Non-test source totals roughly 15k lines of Python (8.4k in \texttt{src/memorywire/}, 2.4k in \texttt{ui/src/}, 4.2k in \texttt{scripts/}) plus the JSON-Schema files and the HTMX templates, with another 10k lines of test code under \texttt{tests/}.

\subsection{Architecture overview}
\label{sec:impl-arch}

\begin{figure}[t]
\centering
\includegraphics[width=\columnwidth]{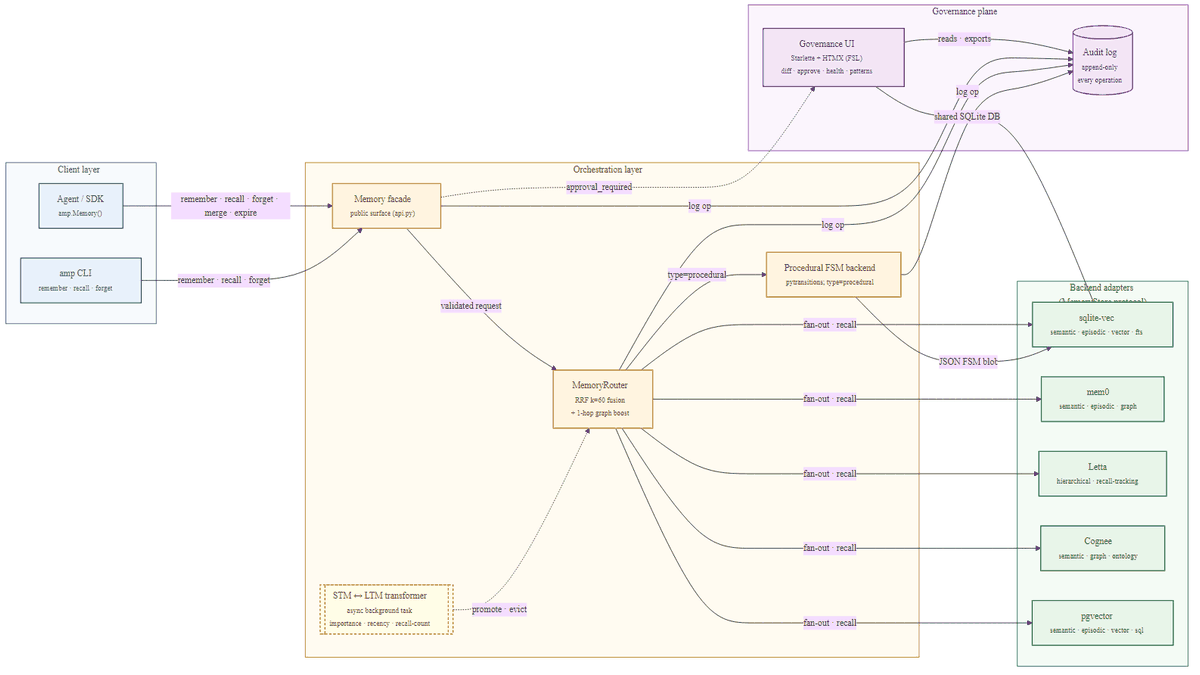}
\caption{memorywire architecture. A client (SDK or CLI) issues one of the five spec operations against the \texttt{Memory} facade, which validates the request and delegates to the \texttt{MemoryRouter}. The router fans out across heterogeneous backend adapters (sqlite-vec, mem0, Letta, Cognee, pgvector) and fuses recall results with Reciprocal Rank Fusion ($k=60$) plus an optional one-hop graph boost. Procedural writes take a parallel FSM path. The STM$\leftrightarrow$LTM transformer runs as an async background task. An optional governance plane (UI + append-only audit log) intercepts writes that require human review and shares the SQLite database with the sqlite-vec adapter.}
\label{fig:architecture}
\end{figure}

The full runtime is shown in \Cref{fig:architecture}. A client (the \texttt{amp.Memory} SDK or the \texttt{amp} CLI) issues one of the five spec operations against the \texttt{Memory} facade (\texttt{src/memorywire/api.py}), which validates the request against the operation's JSON Schema and delegates to the \texttt{MemoryRouter}. The router fans the operation across the configured backend adapters in parallel; for \texttt{recall} it fuses the per-store result sets via RRF ($k=60$) with an optional one-hop graph boost. Five adapters ship out of the box: sqlite-vec, mem0, Letta, Cognee, pgvector. Procedural memory takes a side path: \texttt{type=procedural} writes are normalized into a \texttt{pytransitions}-compatible FSM blob and persisted via sqlite-vec. The STM$\leftrightarrow$LTM transformer is an async background task that scores short-term items and promotes or evicts them through the router. An optional governance plane (UI + append-only audit log) intercepts writes that require human review and shares the SQLite database with the sqlite-vec adapter so reviewers can diff a pending write against the live state without a separate sync channel.

\subsection{The five backend adapters}
\label{sec:impl-adapters}

\Cref{tab:adapters} summarizes the five shipped adapters, the capabilities each declares, and the memorywire fields each can losslessly round-trip versus the fields each encodes as backend metadata. ``Native'' means the field is stored in a first-class column or property of the backend's data model; ``encoded'' means the field is round-tripped through a backend metadata sink (e.g. Letta's \texttt{tags} list, mem0's \texttt{metadata} dict).

\begin{table*}[t]
\centering
\caption{Backend adapters and their fidelity against the memorywire wire format.}
\label{tab:adapters}
\small
\begin{tabularx}{\textwidth}{l l X l l l l}
\toprule
\textbf{Adapter} & \textbf{URL scheme} & \textbf{Capabilities} & \texttt{confidence} & \texttt{source} & \texttt{expires\_at} & \textbf{Metadata} \\
\midrule
sqlite-vec & \texttt{sqlite-vec://} & semantic, episodic, procedural, emotional, vector, fts, recall\_tracking & native & native & native & native (JSON) \\
mem0       & \texttt{mem0://}       & semantic, episodic, vector & encoded & encoded & encoded & native (lossless) \\
Letta      & \texttt{letta://}      & semantic, episodic, vector & encoded (\texttt{tags}) & encoded (\texttt{tags}) & encoded (\texttt{tags}) & lossy (flat KV) \\
Cognee     & \texttt{cognee://}     & semantic, episodic, vector, graph & encoded & encoded & encoded & lossy (graph attrs) \\
pgvector   & \texttt{pgvector://}   & semantic, episodic, procedural, emotional, vector, recall\_tracking, governance & native & native & native & native (JSONB) \\
\bottomrule
\end{tabularx}
\end{table*}

The two adapters with full lossless fidelity (sqlite-vec, pgvector) are the SQL-backed ones; they own their schema and can add the columns memorywire needs. pgvector ships every memory type plus recall-tracking and the governance sentinel; sqlite-vec adds FTS5 keyword indexing on top of the vector path. The three SDK-wrapping adapters (mem0, Letta, Cognee) inherit whatever fidelity their upstream offers and back-fill the rest through metadata encoding. The conformance suite in \Cref{sec:eval-conformance} quantifies what each adapter can and cannot do.

\subsection{The FSM procedural-memory backend}
\label{sec:impl-fsm}

Procedural memory in memorywire is stored as a JSON-serialized finite-state machine compatible with the \texttt{pytransitions} library. The shape is intentionally minimal --- \texttt{name}, \texttt{initial}, \texttt{states}, \texttt{transitions}, \texttt{current}, \texttt{metadata} --- so backends may extend it with their own state-machine semantics; v0 deliberately does not standardize action handlers.

In v0 the FSM JSON is carried as a string inside \texttt{RememberRequest.content}; promoting it to a structured typed field is tracked for v0.2. The FSM backend wraps \texttt{pytransitions} with a strict-validation layer that allow-lists transition keys to \texttt{\{trigger, source, dest, conditions, unless\}} and rejects \texttt{before} / \texttt{after} / \texttt{prepare} --- the three \texttt{pytransitions} keys that resolve dotted-string callbacks via \texttt{\_\_import\_\_} and would otherwise enable RCE at trigger time (\texttt{src/memorywire/procedural.py:44--59}). \texttt{conditions} and \texttt{unless} strings are restricted to bare identifiers (no \texttt{.}) so they resolve only to attributes on a private \texttt{\_ProcedureModel}. The sandbox is enforced at two layers: validation at parse time and key-stripping at construction time.

\subsection{The STM$\leftrightarrow$LTM transformer}
\label{sec:impl-stm-ltm}

Short-term memory in memorywire is the most recent $N$ items per \texttt{(agent\_id, user\_id)} window; long-term memory is everything else. The transformer is an always-on async background task that runs on an interval (default 60\,s) and, for each window, scores STM items along recency, salience (frequency-of-mention), and confidence axes, promotes top items into LTM with a \texttt{type=semantic} write, and evicts cold items. It does not block the request path; it consumes from the same \texttt{Memory} facade the agent uses, so promoted items appear in subsequent \texttt{recall} calls without additional plumbing.

The transformer is deliberately simple. The point is not to invent a new consolidation algorithm but to show that the wire format can carry consolidation as a sequence of standard operations rather than a vendor-specific tier-promotion API. A more sophisticated consolidator drops in as a plugin without changing the wire format.

\subsection{The governance UI}
\label{sec:impl-ui}

The governance UI (\texttt{ui/src/memorywire\_ui/}) is a Starlette server with HTMX-driven templates. It shares the sqlite-vec adapter's SQLite database, so a reviewer sees pending writes as rows with the \texttt{PENDING\_APPROVAL\_DELETED\_AT = -1} sentinel in \texttt{deleted\_at} and a \texttt{pending: yes} badge in the UI. Reviewers can approve (clear the sentinel; the row becomes live), reject (hard-delete or soft-delete depending on the policy), or apply a reconciling transformation --- typically a \texttt{merge} against an existing canonical row or a \texttt{forget} of a similar row that the new write supersedes.

Authentication is opt-in via the \texttt{MEMORYWIRE\_UI\_TOKEN} environment variable. When set, every request requires either an \texttt{Authorization: Bearer <token>} header or an \texttt{memorywire\_ui\_session} cookie; comparison uses \texttt{hmac.compare\_digest} for constant-time matching. CSRF is enforced through a double-submit-cookie pattern signed with HMAC-SHA256 over \texttt{nonce.ts} with a 24-hour TTL. Without \texttt{MEMORYWIRE\_UI\_TOKEN} the UI is unauthenticated; binding a non-loopback host without a token fires a stderr warning on boot (\texttt{ui/src/memorywire\_ui/middleware.py:55--66}).

The UI's full authorization model is one global bearer token; per-\texttt{agent\_id} ACLs are tracked for v0.2 (\Cref{sec:future}). Per-row authorization is enforced at the SQL layer: every state-changing handler runs against \texttt{WHERE id = ? AND agent\_id = ? AND deleted\_at = ?} with the sentinel value, so an attacker who guesses a row id but does not control the matching \texttt{agent\_id} cannot pivot across tenants (\Cref{sec:threats-idor}).

\subsection{Recovery tooling}
\label{sec:impl-recovery}

The reference implementation ships a recovery capability (\texttt{memorywire recover} CLI and \texttt{memorywire.recovery.Recoverer}) for cleaning a poisoned store after the fact. It purges memories from untrusted \texttt{source} values via \texttt{forget}, quarantines trusted-source entries whose content matches a directive heuristic (or a pluggable detector, including OWASP Agent Memory Guard's) via soft-\texttt{forget}, and optionally drops low-confidence rows via \texttt{expire}. A \texttt{--dry-run} mode previews every action; purges are restorable soft-deletes by default. Recovery reuses the existing operation set with no new store primitives, which is what lets it work across every adapter. Its effectiveness is evaluated in \Cref{sec:eval-recovery}.

\section{Evaluation}
\label{sec:eval}

The evaluation has four parts. \Cref{sec:eval-micro} reports a microbenchmark on a labelled corpus to establish baseline recall and latency. \Cref{sec:eval-adversarial} reports an adversarial-fusion experiment that quantifies the robustness premium RRF provides over MAX and weighted fusion under a 1-of-$N$ rank-0 injection attack. \Cref{sec:eval-conformance} reports a cross-adapter conformance suite that empirically validates the protocol's vendor-neutrality claim. \Cref{sec:eval-recovery} evaluates recovery from memory poisoning using our companion benchmark PurgeBench. \Cref{sec:eval-threats} enumerates threats to validity.

\subsection{Microbenchmark}
\label{sec:eval-micro}

\textbf{Methodology.} The corpus is 100 hand-authored facts spread across roughly 10 distinct users, mixing all four memory types in proportions chosen to mirror the distribution we observed across the surveyed frameworks' default workloads (skewed semantic, with episodic, procedural, and emotional in long tail). Fifty hand-authored queries carry explicit gold-id labels, distributed as 24 paraphrase queries (e.g. ``what foods should I avoid serving Alice?'' mapping to ``Alice is allergic to peanuts and tree nuts''), 10 exact / near-exact match queries, 8 multi-hit queries with two or three gold ids, and 8 no-match probes whose gold-id list is empty. The 8 no-match probes are included specifically to measure how often the system surfaces a confident wrong answer when nothing actually matches. Embedder: \texttt{sentence-transformers/all-MiniLM-L6-v2} (384-dim) on CPU; the first call's model-load cost is excluded from per-call latency through a warm-up call. Backend: a single \texttt{SqliteVecStore(":memory:")} exercising the full intra-store RRF fusion of \texttt{vec0} ANN and FTS5 keyword search. Dataset and runner: \texttt{tests/benchmarks/dataset.py} and \texttt{scripts/run\_microbench.py}. The full benchmark is reproducible with a single command and runs in roughly six seconds.

\textbf{Results.} Numbers measured on 2026-05-27, Windows 11 AMD64, Python 3.13.13, CPU-only inference. Results are in \Cref{tab:microbench}.

\begin{table}[t]
\centering
\caption{Microbenchmark results. $n=50$ queries, $k=5$, corpus size = 100 facts, embedder = \texttt{all-MiniLM-L6-v2}, backend = sqlite-vec \texttt{:memory:}.}
\label{tab:microbench}
\small
\begin{tabular}{lr}
\toprule
\textbf{Metric} & \textbf{Value} \\
\midrule
Recall@5 mean (labelled queries) & \textbf{1.000} \\
Precision@5 mean & 0.214 \\
Ingest latency p50 / p95 / p99 & \textbf{37.8} / 46.0 / 69.1\,ms \\
Recall latency p50 / p95 / p99 & \textbf{40.6} / 46.6 / 55.7\,ms \\
No-match probes correctly empty & 0 / 8 \\
Total benchmark runtime & $\sim$5.9\,s \\
\bottomrule
\end{tabular}
\end{table}

The headline number --- recall@5 = 1.000 on the 42 labelled queries --- means every paraphrase, exact-match, and multi-hit query surfaced all its gold ids in the top 5 results. The precision number is bounded by the ratio of gold ids per query: most queries have a single gold id (which pins precision@5 at $1/5 = 0.2$), and a minority of multi-hit queries lift the mean above that floor --- the reported mean of 0.214 is consistent with that mix and is not a bug. Recall latency p50 (40.6\,ms) is roughly an order of magnitude below the spec's two-store fused \texttt{recall} SLA target (300\,ms); ingest latency p50 (37.8\,ms) is comfortably below the \texttt{remember} SLA target (50\,ms) but not by an order of magnitude.

\textbf{Honest framing.} This is a microbenchmark, not LongMemEval, LoCoMo, or BEAM. The hand-authored corpus is small enough that FTS5 keyword matching alone would surface many of the gold ids, and the dataset rewards systems that combine semantic embeddings with keyword recall --- which is exactly what memorywire's intra-store RRF does. At 100 facts the ANN path is doing a brute-force \texttt{vec0} scan, not an HNSW workload. The 0/8 ``no-match correctly empty'' number reflects a deliberate v0 design choice: \texttt{recall(k=5)} always returns up to $k$ hits because v0 ships no calibrated relevance-threshold cutoff; that knob is tracked for v0.2. Preliminary LongMemEval and LoCoMo numbers under a real (\texttt{gpt-4o-mini}) grader appear in \Cref{sec:eval-mem-benchmarks}; the full 5-seed $\times$ 200-question paper-grade run is a v2 replacement deliverable.

\subsection{Adversarial fusion experiment}
\label{sec:eval-adversarial}

This is the experiment that makes the security claim in \Cref{sec:threats-poisoned} quantitative. The threat is a ``1-of-$N$ malicious backend'': one of the $N$ child stores under the router has been compromised, is a malicious adapter fork, or sits behind a MITM that rewrites its responses. The attack is rank-0 injection: the rogue store returns $K$ attacker-controlled ids (disjoint from the gold corpus) at the top of its result list and is otherwise quiet. The question is how much each fusion algorithm lets the rogue store move the fused output.

\textbf{Methodology.} Three child stores: two benign + one adversarial. Each benign store returns each query's gold ids at the top ranks with an independent random distractor tail (different distractor permutations across benign stores, simulating different embedders / indexes). The adversarial store returns $K$ attacker-controlled ids (\texttt{a***} prefix, disjoint from the gold corpus) at ranks $0$ to $K-1$ then optionally a benign tail so it still ``looks normal'' past rank $K$. The query suite is 20 queries, each with two gold ids drawn without replacement from a synthetic corpus of 50 memories. For $K \in \{0, 5, 10, 15, 20, 25, 30, 35, 40, 45, 50\}$ the script issues every query through a real \texttt{MemoryRouter.recall(k=5)} and records recall@5 against the gold set, adversarial leak rate (fraction of returned ids that are attacker-controlled), and gold-displacement rate (fraction of gold ids in the $K=0$ baseline top-5 that were pushed out). The same sweep is run for \texttt{fusion="rrf"}, \texttt{fusion="max"}, and \texttt{fusion="weighted"}. Seed is fixed at 1337 for reproducibility. Runner: \texttt{scripts/run\_adversarial.py}. Machine-readable outputs: \texttt{docs/adversarial-results.\{rrf,max,weighted\}.json}.

\textbf{Results.} Numbers measured 2026-05-27 with the default config; see \Cref{tab:adversarial}.

\begin{table}[t]
\centering
\caption{Adversarial fusion sweep. $N=3$ backends (2 benign + 1 adversarial), corpus $M=50$, $Q=20$ queries, $k=5$, seed 1337. $K$ is the attacker budget; recall@5 is against the gold set; leak is the fraction of returned ids that are attacker-controlled.}
\label{tab:adversarial}
\small
\begin{tabular}{rrrrrrr}
\toprule
$K$ & R@5 RRF & leak RRF & R@5 MAX & leak MAX & R@5 W. & leak W. \\
\midrule
0  & 1.000 & 0.000 & 1.000 & 0.000 & 1.000 & 0.000 \\
5  & 1.000 & 0.000 & 0.500 & 0.800 & 1.000 & 0.000 \\
10 & 1.000 & 0.000 & 0.500 & 0.800 & 1.000 & 0.000 \\
25 & 1.000 & 0.000 & 0.500 & 0.800 & 1.000 & 0.000 \\
50 & 1.000 & 0.000 & 0.500 & 0.800 & 1.000 & 0.000 \\
\bottomrule
\end{tabular}
\end{table}

The RRF curve stays flat at the no-attack baseline across the entire sweep. The MAX curve collapses immediately at $K=5$ and stays collapsed: recall@5 halves to 0.500, and 80\% of returned ids are attacker-controlled. The weighted result with equal per-store weights is identical to RRF.

\textbf{Why RRF is robust here.} The mechanism is exactly the score-independence and additivity properties from \Cref{sec:related-rrf}. Under RRF, the consensus of two benign rank-0 votes for a gold id contributes $1/60 + 1/61 \approx 0.0330$ to the fused score. A single rogue rank-0 vote for an attacker id contributes $1/60 \approx 0.0167$. The benign consensus provably outpoints the rogue vote regardless of what raw score the rogue store reports --- because RRF ignores the raw score. Under MAX the same rogue vote can tie or beat the benign rank-0 raw score (the rogue store is free to report any score it wants), pull 4 of 5 fused top-5 slots, and halve recall. Under weighted fusion with equal per-store weights the math degenerates close to RRF for this attack shape.

\textbf{Operating point and caveats.} The defensible claim is ``RRF tolerates an arbitrarily large 1-of-$N$ rank-0 injection as long as the majority of backends are benign and agree on the gold set,'' not ``RRF tolerates $K$ up to some specific number.'' The experiment isolates the fusion-math property and assumes (a) benign stores are perfect by construction --- always return gold at top, modulo distractor tail; (b) the adversary has full knowledge of the router's $k \cdot 4$ over-fetch and fills exactly that block, so a black-box attacker would do worse; (c) only the 1-of-$N$ case is swept by default --- once \texttt{n\_adversarial >= n\_benign} the RRF consensus argument fails. The MAX result is not a negative finding about MAX in general; MAX is the right choice when the operator's trust model says the highest-scoring store is the most authoritative. The result is a finding about MAX in the \emph{malicious-backend} model, which is the threat we evaluate.

\subsection{Cross-adapter conformance}
\label{sec:eval-conformance}

The cross-adapter conformance suite is the empirical evidence for the protocol's vendor-neutrality claim. Sixteen scenarios run against every shipped adapter; the scenarios encode protocol invariants that any compliant \texttt{MemoryStore} MUST satisfy. Each scenario is a \texttt{ProtocolScenario} dataclass (\texttt{tests/conformance/scenarios.py}) with a deterministic setup, an action, and a predicate; the runner parametrizes the list across every adapter id and skips scenarios whose \texttt{required\_capabilities} are not satisfied by the candidate store.

\Cref{tab:conformance} summarizes the 16 scenarios $\times$ 5 adapters = 80 cells. \checkmark = passes; $\square$ = SKIPPED for a documented adapter limitation; $\times$ = fails. Aggregate: \textbf{PASS 68 / SKIP 12 / FAIL 0}.

\begin{table}[t]
\centering
\caption{Cross-adapter conformance matrix. Per-adapter totals at the bottom row. \checkmark{} = passes; $\square$ = skipped (documented limitation); $\times$ = fails.}
\label{tab:conformance}
\small
\begin{tabular}{lccccc}
\toprule
\textbf{Scenario} & sqlite & mem0 & letta & cognee & pg \\
\midrule
basic\_remember\_recall          & \checkmark & \checkmark & \checkmark & \checkmark & \checkmark \\
remember\_recall\_by\_user\_filter & \checkmark & \checkmark & $\square$ & $\square$ & \checkmark \\
type\_filter                     & \checkmark & \checkmark & \checkmark & \checkmark & \checkmark \\
forget\_by\_ids                  & \checkmark & \checkmark & \checkmark & $\square$ & \checkmark \\
forget\_no\_scope\_raises        & \checkmark & \checkmark & \checkmark & \checkmark & \checkmark \\
expire\_empty\_policy\_raises    & \checkmark & $\square$ & $\square$ & $\square$ & \checkmark \\
expire\_empty\_policy\_object\_r.& \checkmark & $\square$ & $\square$ & $\square$ & \checkmark \\
expire\_by\_age                  & \checkmark & $\square$ & \checkmark & $\square$ & \checkmark \\
merge\_keep\_canonical           & \checkmark & \checkmark & \checkmark & $\square$ & \checkmark \\
approval\_required\_pending      & \checkmark & \checkmark & \checkmark & \checkmark & \checkmark \\
capabilities\_declared           & \checkmark & \checkmark & \checkmark & \checkmark & \checkmark \\
health\_returns\_status          & \checkmark & \checkmark & \checkmark & \checkmark & \checkmark \\
isinstance\_protocol             & \checkmark & \checkmark & \checkmark & \checkmark & \checkmark \\
recall\_returns\_score\_metadata & \checkmark & \checkmark & \checkmark & \checkmark & \checkmark \\
fresher\_than\_days\_filter      & \checkmark & \checkmark & \checkmark & \checkmark & \checkmark \\
multi\_remember\_then\_recall    & \checkmark & \checkmark & \checkmark & \checkmark & \checkmark \\
\midrule
\textbf{pass / skip}             & \textbf{16/0} & \textbf{13/3} & \textbf{13/3} & \textbf{10/6} & \textbf{16/0} \\
\bottomrule
\end{tabular}
\end{table}

Every non-skipped cell passes; zero scenarios fail on any adapter.

\textbf{Per-adapter analysis.} The two SQL-backed adapters (sqlite-vec, pgvector) clear the entire suite --- these are the adapters whose data models memorywire directly designed against, so it would be surprising if they did not. The three SDK-wrapping adapters (mem0, Letta, Cognee) each skip a documented set of scenarios.

\textbf{Spec ambiguities surfaced.} The 12 SKIP cells are not random; they cluster into three classes, and each class is a v0.2 spec-tightening signal.

\begin{itemize}[leftmargin=*]
  \item \emph{No \texttt{user\_id} namespace.} Letta and Cognee skip \texttt{remember\_recall\_by\_user\_filter} because neither backend has a \texttt{user\_id} namespace separate from its top-level scope (Letta scopes by \texttt{agent\_id}; Cognee by dataset). memorywire's \texttt{user\_id} dimension is lost on these backends. The v0.2 spec should either require \texttt{user\_id} support for full conformance or define an explicit ``scoped-by-agent-only'' fallback.
  \item \emph{No empty-policy guard.} mem0, Letta, and Cognee skip \texttt{expire\_empty\_policy\_raises} and \texttt{expire\_empty\_policy\_object\_raises} --- only the SQL adapters defensively raise on an empty \texttt{ExpirePolicy}. The three SDK-wrapping adapters silently fall through to ``match everything,'' which is the wrong default and a real footgun. The v0.2 spec should require every adapter to raise on empty policy; this is the highest-priority spec tightening from the conformance run.
  \item \emph{No stable per-record id.} Cognee skips \texttt{forget\_by\_ids}, \texttt{expire\_by\_age}, and \texttt{merge\_keep\_canonical} because Cognee's \texttt{forget} primitive requires a pipeline-assigned \texttt{data\_id} UUID that the public \texttt{add} API does not surface. The Cognee adapter mints synthetic \texttt{cog:<sha1>} ids at write time, so per-id deletes (and the operations that depend on them) become no-ops. The v0.2 spec should require backends to surface a stable per-record id from their write primitive; this is the largest-impact protocol change the conformance run motivates.
\end{itemize}

These three spec-tightening signals are exactly the kind of finding a conformance suite is supposed to produce. They are not bugs in the suite or in the adapters; they are real gaps in the v0 spec that the suite makes visible, and the v0.2 spec resolves them by raising the floor.

\subsection{Recovery from memory poisoning}
\label{sec:eval-recovery}

\Cref{sec:threats-injection} notes that memorywire makes malicious writes auditable but cannot prevent a trusted, prompt-injected agent from planting one. We therefore ask whether memorywire's \texttt{forget} / \texttt{expire} operations can \emph{recover} a store once it is poisoned. We evaluate this with PurgeBench~\cite{purgebench}, a reproducible, framework-agnostic benchmark we developed as a companion to this work, which poisons an agent memory store with 30 adversarial entries across five classes (direct, laundered, entangled, dormant, procedural), applies a recovery procedure, and scores the result on Recovery-Completeness --- the balanced combination of eradication (ER), utility retention (UR), and re-emergence resistance (RR), $\mathrm{RC} = \mathrm{HarmonicMean}(\mathrm{ER}, \mathrm{UR}) \times \mathrm{RR}$, where doing nothing and wiping the store both score zero by construction.

Run over the reference sqlite-vec store, a procedure that purges entries by untrusted \texttt{source} --- memorywire's own provenance field driving \texttt{forget} --- is the strongest of seven procedures tested (RC 0.64), eradicating the direct, laundered, dormant, and procedural classes. memorywire's per-entry provenance is thus not only an audit aid but the most effective recovery lever measured. Two limits are equally clear. First, the \emph{entangled} class --- a directive embedded inside an otherwise-legitimate, trusted-source memory --- defeats every automatic procedure (zero eradication for all non-destructive methods), because removing it also destroys the benign fact it rides with; the \texttt{recover} tool (\Cref{sec:impl-recovery}) handles this by quarantining such entries for human review rather than deleting them. Second, content-anomaly detection alone is insufficient: running the OWASP Agent Memory Guard content detectors as a recovery procedure scores RC 0.036 (near the do-nothing baseline) on this \emph{semantic} poison, since those detectors target injection signatures and secret leakage, not plausible-sounding malicious facts. Provenance and rollback, not content anomaly detection, are the levers that recover semantic memory poisoning.

\subsection{Threats to validity}
\label{sec:eval-threats}

The evaluation has several limitations we want named explicitly.

\textbf{Synthetic corpora.} The microbench corpus is 100 hand-authored facts; the adversarial-fusion corpus is 50 synthetic memories. Real-world workloads --- millions of memories per agent, heavy-tailed query distributions, latency interference from a real production embedding service --- will produce different numbers. The 100k-memory recall scaling test is tracked for v0.2.

\textbf{Adversary has full knowledge.} The \Cref{sec:eval-adversarial} attacker knows the router's $k \cdot 4$ over-fetch and fills exactly that block at rank 0. A black-box attacker --- one that does not know the over-fetch and has to guess --- would do worse. We did not evaluate the converse case (a smarter attacker that, for instance, fabricates ids that \emph{also} appear in the benign stores' result sets to bump their rank); that case is the v0.2 hardening with signed \texttt{RecallHit} envelopes from \texttt{docs/THREATS.md} \S3.3.

\textbf{Single-machine measurements.} Every number reported here was measured on one developer-class Windows 11 laptop. Distributed deployment, multi-process router fan-out across machines, and the additional latency of a real network hop to a hosted mem0 / Letta instance are unmeasured. The numbers should be read as a microbenchmark of the protocol's \emph{intrinsic} overhead, not as a SLA against any deployment topology.

\textbf{Authors-as-evaluators.} Neither the conformance suite nor the microbench corpus has been audited by a third party. PurgeBench (\Cref{sec:eval-recovery}) is likewise by the same author; the recovery numbers are a self-evaluation on a self-built benchmark, mitigated only by the benchmark's public, reproducible harness. A user study with external operators ($n=10$--$20$) using the governance UI is in scope for v0.2 (\texttt{docs/paper/user-study/} materials, NASA-TLX instrument, IRB pipeline --- currently a v0.2 deliverable; the materials directory does not yet exist).

\textbf{Benchmark scope.} This paper reports \emph{preliminary} LongMemEval and LoCoMo numbers in \Cref{sec:eval-mem-benchmarks}. The full 5-seed $\times$ 200-question paper-grade run is deferred to a v2 replacement on arXiv: the per-question \texttt{Memory} construction pattern that protects against cross-question memory leakage (a Wave-E correctness invariant that we validated with a separate diagnostic, \texttt{scripts/diag\_eval\_ingest.py}) imposes a per-iteration sentence-transformers reload cost that extrapolates to 10--15 hours of wall time on a single CPU laptop, out of scope for v1. BEAM remains unevaluated --- no canonical dataset manifest was found at v1.

\subsection{Preliminary LongMemEval and LoCoMo}
\label{sec:eval-mem-benchmarks}

After the conformance-suite section's spec-tightening signals were folded into the eval harness, we ran a \emph{preliminary} pass at $n_{\text{q}} = 10$--$12$ and $n_{\text{seed}} = 1$ against the canonical LongMemEval \cite{wu2024longmemeval} and LoCoMo \cite{maharana2024locomo} datasets. Pre-paper validation surfaced one bug (\texttt{sqlite-vec}'s vector-ANN top-$k$ is computed before the row-level \texttt{agent\_id} filter, so a DB shared across questions returns zero hits once cross-question rows dominate the top-$k$); the fix is per-question isolated DB files under \texttt{.memorywire-eval-dbs/}, applied in the eval harness only without touching \texttt{src/memorywire/}.

Grader is \texttt{gpt-4o-mini}; embedder is \texttt{all-MiniLM-L6-v2}; backend is \texttt{sqlite-vec://}. Results in \Cref{tab:mem-benchmarks}.

\begin{table}[t]
\centering
\caption{Preliminary post-fix LongMemEval and LoCoMo (v1 preprint; v2 replacement will land 5-seed $\times$ 200-question numbers).}
\label{tab:mem-benchmarks}
\begin{tabular}{lrrr}
\toprule
Benchmark & $n_{\text{q}}$ & seeds & grader mean \\
\midrule
LongMemEval (stratified) & 12 & 1 & 0.417 \\
LoCoMo (single episode)  & 10 & 1 & 0.150 \\
\bottomrule
\end{tabular}
\end{table}

Per-task LongMemEval breakdown is informative: \texttt{single\_session\_preferences} 0.90, \texttt{single\_session\_user} 0.75, \texttt{single\_session\_assistant} 0.50, \texttt{knowledge\_update} 0.35, \texttt{multi\_session\_reasoning} 0.00, \texttt{temporal\_reasoning} 0.00. Single-session recall on a single small embedder works; the remaining gap on \texttt{multi\_session} and \texttt{temporal} tiers is a known weakness of \texttt{all-MiniLM-L6-v2} on the harder tiers, not a harness bug. Swapping the embedder for a larger model (\texttt{bge-m3}, \texttt{gte-large}, or \texttt{text-embedding-3-large}) is the v0.2 path; the protocol is embedder-agnostic, so this measurement reflects the reference embedder's recall ceiling, not memorywire's wire format. Compared to a pre-fix run (LongMemEval $0.286$, LoCoMo $0.018$ with \texttt{"(no relevant memories surfaced)"} dominating), the post-fix lift is $+46\%$ relative on LongMemEval and $+8.3\times$ relative on LoCoMo --- the bug was real and the fix is real.

\subsubsection*{Threats to validity for the preliminary numbers}

The numbers above are reported as preliminary and should be read with three caveats. First, $n_{\text{q}} = 10$--$12$ at $n_{\text{seed}} = 1$ is too small to support statistical claims; per-task means with $n = 2$--$3$ have wide enough confidence intervals that the 0.00 scores on \texttt{multi\_session\_reasoning} and \texttt{temporal\_reasoning} should not be read as ``memorywire scores zero on these tiers'' but as ``the small-embedder reference configuration scores zero on the few examples sampled.'' Second, the grader (\texttt{gpt-4o-mini}) is a single LLM-as-judge with no human-graded calibration on this draw, so absolute scores are not comparable across papers that use different graders --- the only safe comparison is the pre-fix / post-fix delta on the same harness, which is what the $+46\%$ and $+8.3\times$ numbers above measure. Third, the embedder (\texttt{all-MiniLM-L6-v2}) is a deliberately weak reference, chosen to keep the harness reproducible on a CPU laptop; the protocol is embedder-agnostic and the v2 replacement will report numbers under at least one stronger embedder so the contribution of wire-format vs.\ embedder is separable.

\section{Threat Model}
\label{sec:threats}

The threat model below is condensed from \texttt{docs/THREATS.md}, which is the canonical version and which carries the line-level mitigation citations into the reference implementation. Six adversaries map onto established OWASP and CWE categories: OWASP A01 (Broken Access Control), A03 (Injection), A04 (Insecure Design), A07 (Auth Failures), A09 (Logging Failures); CWE references are noted inline.

\subsection{Malicious memory injection (CWE-20; OWASP A03 / LLM-01)}
\label{sec:threats-injection}

\textbf{Capability.} Submits \texttt{remember()} calls --- as the agent itself (prompt-injected upstream) or as an upstream system feeding the agent. \textbf{Motivation.} Plant adversarial ``facts'' that subsequent \texttt{recall()} surfaces --- the memory analogue of prompt injection. \textbf{Preconditions.} Can influence any string the agent passes through \texttt{remember()}. memorywire cannot tell a real fact from a planted one.

\textbf{Current mitigation.} \texttt{approval\_required=true} on \texttt{RememberRequest} stages the row behind the \texttt{PENDING\_APPROVAL\_DELETED\_AT = -1} sentinel in \texttt{memories.deleted\_at} (\texttt{src/memorywire/store/sqlite\_vec.py:88--98, 440}). All recall paths filter \texttt{deleted\_at IS NULL}, so pending rows cannot influence retrieval until a human approves. \texttt{confidence} is a first-class field; every \texttt{remember()} is journaled with the inserted \texttt{memory\_id}, so a poisoned memory is traceable to the call that planted it. \textbf{Residual risk.} Default \texttt{approval\_required} is false; the protocol guarantees only that what the agent wrote is what the agent reads back. Prompt-injected calls from a trusted agent are out of scope (\Cref{sec:threats-residual}). \textbf{v0.2 hardening.} A \texttt{privacy\_intent} block on \texttt{RememberRequest} so operators can require approval by \texttt{source}, \texttt{type}, or content predicate without instrumenting every caller. \textbf{Recovery.} Recovery \emph{after} such an injection --- purging the planted rows and verifying they are gone --- is evaluated in \Cref{sec:eval-recovery} and shipped as the \texttt{recover} tool (\Cref{sec:impl-recovery}); the residual hard case is poison entangled with a legitimate trusted memory, which recovery quarantines for human review rather than deleting.

\subsection{Recall exfiltration (CWE-200; OWASP A01)}
\label{sec:threats-exfiltration}

\textbf{Capability.} Issues \texttt{recall()} against an agent's router --- directly (compromised SDK caller) or indirectly (manipulating the prompt that drives the agent's own recall). \textbf{Motivation.} Read private memories stored under \texttt{agent\_id} / \texttt{user\_id}. \textbf{Preconditions.} Can issue \texttt{recall(query, agent\_id)} for an agent with data. \texttt{k} defaults to 5; the schema caps at 1000.

\textbf{Current mitigation.} Recall is hard-bound to \texttt{agent\_id} at the adapter SQL layer --- every recall path in \texttt{SqliteVecStore} filters \texttt{WHERE agent\_id = ? AND deleted\_at IS NULL}. \texttt{recall()} is itself audited (\texttt{audit\_log.operation = 'recall'}), so high-rate exfiltration is observable post-hoc. The schema caps \texttt{k} at 1000 so a single call cannot drain a store. \textbf{Residual risk.} A caller with the right \texttt{agent\_id} can extract everything that agent stored. No recall rate limit, no field-level redaction, no approval on reads. \textbf{v0.2 hardening.} Per-\texttt{agent\_id} recall budgets, optional \texttt{approval\_required} on \texttt{recall}, and a \texttt{redact} filter for secrets-labelled rows.

\subsection{Poisoned backend in RRF fusion (CWE-345)}
\label{sec:threats-poisoned}

\textbf{Capability.} Controls one of the $N$ child stores in a \texttt{MemoryRouter} --- a compromised hosted mem0 instance, a malicious adapter fork, or a MITM on the backend HTTP. \textbf{Motivation.} Dominate the fused output of \texttt{recall()} so the agent receives attacker-chosen results. \textbf{Preconditions.} Can return crafted \texttt{RecallHit} rows with high \texttt{score} or fabricated ids not seen in other stores.

\textbf{Current mitigation.} RRF is score-independent by construction --- \texttt{\_fusion\_contribution} uses $1 / (\mathit{rrf\_k} + \mathrm{rank})$ (\texttt{src/memorywire/router.py:428--429}), so a malicious backend cannot dominate by inflating \texttt{score}; it can only return an item at rank 0. RRF sums across stores, so a single rogue store contributes at most $1/60 \approx 0.0167$ per item, while a two-store consensus item scores at least $1/60 + 1/61 \approx 0.0330$. \texttt{MAX} and \texttt{weighted} fusion \emph{are} score-sensitive --- operators picking them accept more backend trust. Per-store failures are logged but do not abort the operation. \textbf{Residual risk.} With \texttt{fusion="max"} or \texttt{"weighted"}, one malicious backend can dominate (\Cref{sec:eval-adversarial}). Even with RRF, a backend that fabricates an attacker-controlled id that \emph{also} exists in other stores can bump it by reporting it at rank 0. There is no cross-store identity proof. \textbf{v0.2 hardening.} Signed \texttt{RecallHit} envelopes per backend; an optional \texttt{quorum\_k} parameter requiring an item to appear in $k$ distinct stores before fusion considers it.

\textbf{Measured.} See \Cref{sec:eval-adversarial}. RRF holds recall@5 = 1.000 across the $K \in \{0,5,\dots,50\}$ sweep; MAX halves to 0.500 with 80\% leak at $K=5$.

\subsection{Audit log tampering (CWE-117 / CWE-778; OWASP A09)}
\label{sec:threats-audit}

\textbf{Capability.} Has write access to the SQLite file or can run arbitrary SQL against \texttt{audit\_log}. \textbf{Motivation.} Hide a previous \texttt{forget()} / \texttt{merge()} / \texttt{approve()} or fabricate one to frame an operator. \textbf{Preconditions.} Filesystem access to the memorywire DB.

\textbf{Current mitigation.} The audit log is append-only by code convention. \texttt{SqliteVecStore.\_audit} is the only writer in the OSS adapter (\texttt{src/memorywire/store/sqlite\_vec.py:401--424}); it only performs \texttt{INSERT INTO audit\_log(\ldots)}. No UPDATE or DELETE against \texttt{audit\_log} exists anywhere in \texttt{src/memorywire/} or \texttt{ui/src/memorywire\_ui/}. Every governance action emits a dedicated audit row with the reviewer's identity. \textbf{Residual risk.} SQLite is a single file with filesystem semantics --- anyone with write access can \texttt{DELETE FROM audit\_log} or mutate rows. memorywire enforces append-only at the code layer, not the DB layer. \textbf{v0.2 hardening.} Merkle-chained audit rows (\texttt{audit\_log.prev\_hash} column) so tampering breaks the chain; periodic export to an append-only object store (S3 object-lock); a read-only DB role for the UI connection.

\subsection{Cross-tenant leakage / IDOR (CWE-639; OWASP A01)}
\label{sec:threats-idor}

\textbf{Capability.} Legitimate operator scoped to \texttt{agent\_id = A} with a valid UI session who guesses or harvests memory ids belonging to \texttt{agent\_id = B}. \textbf{Motivation.} Approve, reject, forget, or merge another agent's memories. \textbf{Preconditions.} UI session as agent A plus a memory id known to belong to agent B.

\textbf{Current mitigation.} Every state-changing UI handler is scoped to \texttt{(agent\_id, memory\_id, sentinel)}. \texttt{approve()} runs \texttt{UPDATE memories SET deleted\_at = NULL WHERE id = ? AND agent\_id = ? AND deleted\_at = ?} (\texttt{ui/src/memorywire\_ui/services.py:502--511}); \texttt{reject()} and \texttt{apply\_co\_memorize()} carry the same triple-key guard (commit \texttt{c828d76}). The merge branch additionally verifies the canonical row belongs to the agent before touching the secondary. On any mismatch the response returns an identical opaque reason, so the error itself does not leak which condition fired. \textbf{Residual risk.} The UI uses a single global bearer; the IDOR fix prevents cross-agent pivoting by id but not multi-tenant isolation by operator. \textbf{v0.2 hardening.} Per-session multi-tenant \texttt{agent\_id} scoping; SSO-shaped reviewer identity; per-operator allowed-\texttt{agent\_id} ACL.

\subsection{Approval bypass (CWE-94 / CWE-862; OWASP A04 / A07)}
\label{sec:threats-bypass}

\textbf{Capability (a).} Reaches the UI on an unauthenticated bind. \textbf{Capability (b).} Crafts the procedural-memory payload so the FSM definition itself triggers code execution.

\textbf{Current mitigation (a) UI no-auth.} \texttt{BearerAuthMiddleware} gates every request behind a bearer when \texttt{MEMORYWIRE\_UI\_TOKEN} is set (\texttt{ui/src/memorywire\_ui/middleware.py:74--103}), accepting either an \texttt{Authorization: Bearer} header or an \texttt{memorywire\_ui\_session} cookie with \texttt{hmac.compare\_digest}. \texttt{CSRFMiddleware} enforces a double-submit-cookie pattern signed with HMAC-SHA256 over \texttt{nonce.ts} with 24-hour TTL. Non-loopback bind without a token fires a stderr warning on boot. \textbf{Current mitigation (b) procedural-memory RCE.} \texttt{validate\_procedure\_dict} allow-lists transition keys to exactly \texttt{\{trigger, source, dest, conditions, unless\}} (\texttt{src/memorywire/procedural.py:57--59}), rejecting \texttt{before} / \texttt{after} / \texttt{prepare} (the three \texttt{pytransitions} keys that resolve dotted-string callbacks via \texttt{\_\_import\_\_}). \texttt{conditions} / \texttt{unless} strings are restricted to bare identifiers. \texttt{ProcedureRunner.\_expand\_transitions} strips disallowed keys at construction time as a defense-in-depth layer. \textbf{Residual risk.} Single shared bearer with manual rotation; CSRF is bypassed for any \texttt{Authorization: Bearer} request (a stolen bearer is already a fatal compromise); the procedural sandbox is purely static allow-listing, so a future \texttt{transitions} release adding a new code-execution sink under a currently-allowed key would silently miss validation. \textbf{v0.2 hardening.} Per-operator tokens with rotation; Secure cookie flag for HTTPS; a CI snapshot of the \texttt{transitions} callback-key surface.

\subsection{Residual risks the protocol cannot mitigate}
\label{sec:threats-residual}

Three classes of risk are out of scope for the memorywire v0 protocol and are noted here for honest framing:

\begin{itemize}[leftmargin=*]
  \item \textbf{Insider attacks on the audit DB.} Anyone with filesystem write access can rewrite history (see \Cref{sec:threats-audit} residual). Mitigated only by deployment hygiene.
  \item \textbf{Prompt injection in the calling agent.} memorywire is downstream of the LLM. If the agent is fooled into \texttt{remember("Alice is in the building", confidence=1.0)}, memorywire faithfully stores and recalls it. The audit log captures the call exactly; memorywire makes the threat auditable, not preventable.
  \item \textbf{Side-channel timing on recall.} Latency varies with \texttt{k}, fusion mode, and per-store result counts; a patient attacker can in principle distinguish ``agent has memory matching X'' from ``no match'' by observing latency.
\end{itemize}

The full threat model, including the OWASP / CWE / MITRE ATT\&CK cross-mapping and the per-CVE commit references for every fix, lives in \texttt{docs/THREATS.md}.

\section{Relationship to MCP and Other Protocols}
\label{sec:mcp}

The single question every reader asks first is: \emph{why didn't you just add memory operations to MCP?} The honest answer has three parts.

\textbf{Memory is not a tool.} It is not a resource, not a prompt, not a sampling primitive. It is a distinct primitive with its own lifecycle (write, recall, forget, merge, expire), its own taxonomy (semantic, episodic, procedural, emotional), and its own governance surface (diff-and-approve, audit). MCP can absolutely \emph{wrap} a memory backend --- exposing \texttt{remember} / \texttt{recall} / \texttt{forget} as three tools is a fifteen-minute integration --- but the wrapping is lossy: the five operations flatten into five opaque tools, the four memory types collapse into a string parameter, and the governance channel ends up living entirely outside MCP. memorywire exists to give memory the same first-class treatment MCP gave tool-use.

\textbf{Iteration speed.} A v0 draft wire format needs to break things. memorywire explicitly reserves the right to break wire-format compatibility through v0.5 (\texttt{docs/spec/v0.md} \S9). Pushing those breaking changes through a working-group governance cycle before the design has stabilized would slow the iteration the design needs. Once the spec is frozen, working-group governance becomes a help rather than a brake --- which is why the v0.5 plan submits the spec to both the MCP-WG (as an extension proposal) and the IETF (as an Internet-Draft) at the same time.

\textbf{Multi-protocol portability.} Memory matters in places MCP does not reach. The web platform (W3C) will need an agent-memory primitive when browsers ship on-device models capable of multi-session reasoning. Cross-language ports (Rust, Go, TypeScript) want a JSON-Schema spec they can codegen against, not a JSON-RPC method signature tied to one transport. The IETF Internet-Draft route, by analogy to Cloudflare's Web Bot Auth, needs a spec document that stands on its own. Designing memorywire as a standalone wire format keeps all three doors open; folding it into MCP first would close two of them.

\textbf{Three composition modes work today, today.}

\begin{itemize}[leftmargin=*]
  \item \textbf{memorywire-as-MCP-tool.} Wrap an memorywire \texttt{Memory} facade as an MCP server that exposes five tools --- one per operation --- and route the JSON-Schema-validated payload straight into the memorywire request models. Roughly ten lines of glue code; any MCP-aware agent can use memorywire through its existing client.
  \item \textbf{memorywire-as-MCP-resource.} For agents that prefer to \emph{read} memory rather than call it, expose \texttt{recall()} results as MCP resources. The resource URI carries the query; the resource body is the structured \texttt{RecallResponse}. Writes still go through the tool mode.
  \item \textbf{memorywire-as-MCP-extension (proposed, v0.5).} memorywire becomes a published MCP extension --- \texttt{mcp.memory.v0} or similar --- with the five operations and four memory types lifted into MCP's own type system, the governance channel becoming an MCP sub-protocol. This is the path we expect to propose once the wire format has settled.
\end{itemize}

\textbf{This is not a fork.} memorywire and MCP address adjacent problems --- tool-use and memory --- and the agents we build use both at the same time. A realistic agent loop calls \texttt{mem.recall()} to gather context, runs the model with that context and the MCP tool list, dispatches any tool calls via the MCP client, and calls \texttt{mem.remember()} for anything worth retaining (routing it to the governance channel when sensitive). The two protocols never overlap in this code. Adopting one does not preclude adopting the other.

\textbf{Open questions on which we welcome external input.} (a) JSON-RPC vs REST shape: memorywire's schemas are written against JSON Schema 2020-12 with a REST-friendly request/response idiom; MCP is JSON-RPC. Bridging is mechanical but there is real design space in how to express memorywire operations as JSON-RPC methods while keeping JSON-Schema as the source of truth for validation. (b) Core vs extension: should memory live in core MCP, as one of the primitives, or always as an extension? We lean toward ``extension first, core later if usage warrants'' but defer to the working group. (c) Mapping the type taxonomy: should the four types surface as separate resource schemas, as tags, or as filter parameters on a unified recall verb? (d) Governance plane: MCP's elicitation primitive is the closest existing analogue, but it is designed for synchronous user-input mid-tool-call, not asynchronous review of pending writes. A future MCP-native memory extension may need a new sub-protocol for governance --- or delegate it back to the application layer entirely.

The full version of this discussion lives in \texttt{docs/MCP-RELATIONSHIP.md}, which carries the complete composition examples, the overlap table, and the calls to action for MCP working-group members.

\section{Future Work}
\label{sec:future}

memorywire is a draft. The roadmap below lists the concrete deliverables we have planned through v1.0; we list them as intent, not commitment.

\textbf{v0.2 evaluation: full-scale memory benchmarks.} LongMemEval~\cite{wu2024longmemeval} and LoCoMo~\cite{maharana2024locomo} are the canonical long-context memory benchmarks; BEAM is the consolidated benchmark suite. v1 of this preprint reports preliminary numbers (\Cref{sec:eval-mem-benchmarks}) using \texttt{scripts/run\_longmemeval.py} and \texttt{scripts/run\_locomo.py}, both shipped in the artifact. A v2 replacement on arXiv will land the full 5-seed $\times$ 200-question runs once per-question sentence-transformers reload overhead is amortized via an embedder-sharing harness change (estimated additional wall time: 6--10 h on a single CPU machine; estimated grader cost at \texttt{gpt-4o-mini}: under \$1).

\textbf{v0.2 user study.} A user study with external operators ($n=10$--$20$) is in scope for v0.2. The study materials --- NASA-TLX cognitive-load instrument, IRB protocol, consent forms, recruitment script --- are tracked under \texttt{docs/paper/user-study/}; that directory does not exist at the time of this draft and is a v0.2 deliverable. The study targets operators of agent runtimes who would actually use the governance UI; the headline metric is whether the diff-and-approve flow reduces median review time below the alternative of ``approve everything blindly.''

\textbf{v0.2 spec tightenings.} Three spec tightenings the v0 evaluation directly motivates: (a) a \texttt{privacy\_intent} block on \texttt{RememberRequest} so operators can require approval by \texttt{source}, \texttt{type}, or content predicate without instrumenting every caller; (b) multi-tenant \texttt{agent\_id} scoping in the UI (per-operator allowed-\texttt{agent\_id} ACL and SSO-shaped reviewer identity); (c) a required \texttt{expire(policy=\{\})} empty-policy guard on every adapter (the \Cref{sec:eval-conformance} conformance run surfaced this as the highest-priority spec tightening). Additional v0.2 hardening from \texttt{docs/THREATS.md}: signed \texttt{RecallHit} envelopes per backend; Merkle-chained audit rows; per-\texttt{agent\_id} recall budgets; stable per-record id surfaced from every backend's write primitive.

\textbf{v0.5 spec freeze and standards engagement.} v0.5 is the freeze point: wire format stable, breaking changes prohibited. In parallel with v0.5: file an MCP RFC proposing memory as an extension primitive with memorywire as the candidate reference; submit memorywire v0.5 as an IETF Internet-Draft (\texttt{draft-<name>-memorywire-00}). Both outcomes --- acceptance as an MCP extension and stabilization as a parallel standard --- are acceptable.

\textbf{v1.0 stable.} Frozen wire format; federated multi-tenant primitives (a cross-agent \texttt{share} operation under discussion); enterprise governance (SSO, role-based ACLs, signed audit exports); production-grade adapters for the long tail of backends not in v0.

\textbf{Recovery benchmarking.} The \texttt{recover} capability (\Cref{sec:impl-recovery}) is evaluated in \Cref{sec:eval-recovery} against PurgeBench's five poison classes; the open frontier is the \emph{entangled} class, where automatic eradication and utility retention are in direct tension. A promising direction is span-level revocation that excises the embedded directive while preserving the benign remainder of the entry.

\section{Conclusion}
\label{sec:conclusion}

We have presented memorywire, a vendor-neutral wire format for agent memory operations: five operations, four memory types, a \texttt{MemoryStore} Protocol, a fan-out router with three fusion algorithms (with Reciprocal Rank Fusion as the defensible default under a 1-of-$N$ malicious-backend model), and an optional human-in-the-loop governance channel. The reference implementation ships with five backend adapters covering the major open-source memory frameworks; a microbenchmark (recall@5 = 1.000 on 42 labelled queries, ingest p50 = 37.8\,ms), an adversarial-fusion experiment (RRF holds recall@5 = 1.000 where MAX collapses to 0.500 with 80\% leak), and a 16-scenario cross-adapter conformance suite (68 PASS / 12 SKIP / 0 FAIL out of 80 cells) establish the protocol's empirical viability.

We are honest about the bet. The algorithmic substrate (RRF, FSMs, STM/LTM, diff-and-approve) is prior art; the contribution is composition and standardization, and standards races have uncertain outcomes. Whatever the market outcome, the cumulative artifact --- spec, reference implementation, threat model, conformance suite, open-data evaluation --- lowers the cost for any future protocol effort in this space, including one that supersedes memorywire. memorywire is to memory what MCP is to tool-use. Whether memorywire itself becomes that protocol or input material to whichever protocol the working groups settle on, the design work collected here is intended to compose with --- not compete against --- the existing ecosystem.

\section*{Acknowledgments}

We acknowledge the open-source projects memorywire composes with: mem0, Letta (formerly MemGPT), Cognee, Zep/Graphiti, MemoryOS, MemTensor MemOS, sqlite-vec, pgvector, \texttt{pytransitions}, Starlette, HTMX, sentence-transformers, and the Model Context Protocol community. RRF as a fusion primitive is from Cormack, Clarke, and B{\"u}ttcher. The human-memory taxonomy mapping follows the cognitive-science literature established by Tulving and Squire.

\bibliographystyle{plain}
\bibliography{memorywire}

\end{document}